\documentclass{article}

\usepackage{arxiv}

\usepackage{graphicx}
\usepackage{dcolumn}
\usepackage{amsmath,amsfonts}  
\usepackage{bm}
\usepackage{color}
\usepackage{colortbl,xcolor}

\DeclareMathOperator*{\argmin}{argmin}

\definecolor{green}{cmyk}{0.75002 0 1 0}
\definecolor{yellow}{cmyk}{0.04 0.3 1 0.02}
\definecolor{orange}{cmyk}{0 0.6 1 0}
\definecolor{blue}{rgb}{ 0    0.3470    0.7410}
\definecolor{red}{rgb}{1 0.2 0}
\definecolor{ltred}{rgb}{1 0.7 0.7}
\definecolor{gray}{rgb}{0.8 0.8 0.8}

\title{
Machine learning in fluid dynamics:\\ A critical assessment
}

\author{
  Kunihiko Taira\\
  Department of Mechanical and Aerospace Engineering\\
  University of California, Los Angeles, California 90095, USA\\
  \texttt{ktaira@seas.ucla.edu} \\
  \And
  Georgios Rigas\\
  Department of Aeronautics\\
  Imperial College London, London, SW7 2AZ, UK\\
  \texttt{g.rigas@imperial.ac.uk} \\
  \And
  Kai Fukami\\
  Department of Aerospace Engineering\\
  Tohoku University, Sendai, Miyagi 980-8579, Japan\\
  \texttt{kfukami1@tohoku.ac.jp} \\
}

\date{\today}

\begin{document}

\maketitle

\begin{abstract}
The fluid dynamics community has increasingly adopted machine learning to analyze, model, predict, and control a wide range of flows. These methods offer powerful computational capabilities for regression, compression, and optimization. In some cases, machine learning has even outperformed traditional approaches. However, many fluid mechanics problems remain beyond the reach of current machine learning techniques.  As the field moves from its current state toward a more mature paradigm, this article offers a critical assessment of the key challenges that must be addressed. Tackling these technical issues will not only deepen our understanding of flow physics but also expand the applicability of machine learning beyond fundamental research.  We also highlight the importance of community-maintained datasets and open-source code repositories to accelerate progress in this area.  Furthermore, the future success of machine learning in fluid dynamics will depend on effective training --- not only for the next generation of researchers but also for established fluid mechanicians adapting to this evolving landscape.  Data-driven fluid dynamics is in its critical transitional state over the next few years to shape its future.  This perspective article aims to spark discussions and encourage collaborative efforts to advance the integration of machine learning in fluid dynamics.
\end{abstract}


\section{Introduction}

Fluid dynamics has three subfields of experimental, theoretical, and computational fluid dynamics.
In the past several years, data-driven fluid dynamics has emerged as a powerful fourth subfield to support the overall endeavor of fluid dynamics by leveraging data from experiments and simulations.  The recent level of interest in data-driven techniques by the research community has been staggering, fueled by the enormous advancements in computational hardware and software developments, which especially have enabled the use of deep learning.  Indeed, it has been amazing to observe the rapid and widespread growth of data-driven fluid dynamics in recent years \cite{Brenner:PRF19,BNK2020}.

In its early stage of data-driven analysis, most studies involved the application of basic data-science techniques to canonical fluid flow problems to analyze and model flow physics.  Such approaches often leveraged linearization and/or space-time decomposition to analyze flows.  However, the advent of modern machine learning, particularly its ability to handle non-convex optimization and nonlinear functional approximations, has enabled the incorporation of nonlinear analysis into data-driven frameworks. This advancement allows researchers to capture complex, nonlinear flow physics and correlations that were previously difficult to resolve using traditional methods.  These efforts have opened new avenues to study, model, and control complex fluid flows without stringent assumptions \cite{Lee:PF97,Milano:JCP02}.  This point is a major strength of machine learning and supports the analysis of fluid flows with complex physics.

Modern machine-learning techniques have been used for a range of fluid mechanics problems in recent years \cite{BNK2020}.  For compression of flow field data, nonlinear machine learning methods can exhibit superb performance compared to traditional approaches \cite{Kramer:AIChE91,fukagata2025compressing}.  The resulting low-dimensional representation of complex flow fields can be used to model their dynamics and perform control \cite{Linot23,Fukami:NC23,Fukami:JFM24}.  Finding nonlinear relationships between input and output variables is another strength of machine learning, enabling field estimation \cite{Manohar:IEEE18,Fukami:NMI21} and super resolution \cite{Fukami:JFM19,kim2021unsupervised,MK2025_JFM}.   Moreover, there are major efforts in using machine learning for the development of turbulence models  \cite{duraisamy2019turbulence,ling2016reynolds,lozano2023machine,oulghelou2024machine,Bae:NC22} and optimizing flow control \cite{Rabault:JFM19}, navigation \cite{Gunnarson:NC21,ZLiu:AIAAJXX}, and design strategies \cite{GEORGIOS}.  These efforts are only a handful of studies that the fluid dynamics community has seen.

\begin{figure}
  \centerline{
  \includegraphics[width=0.8\textwidth]{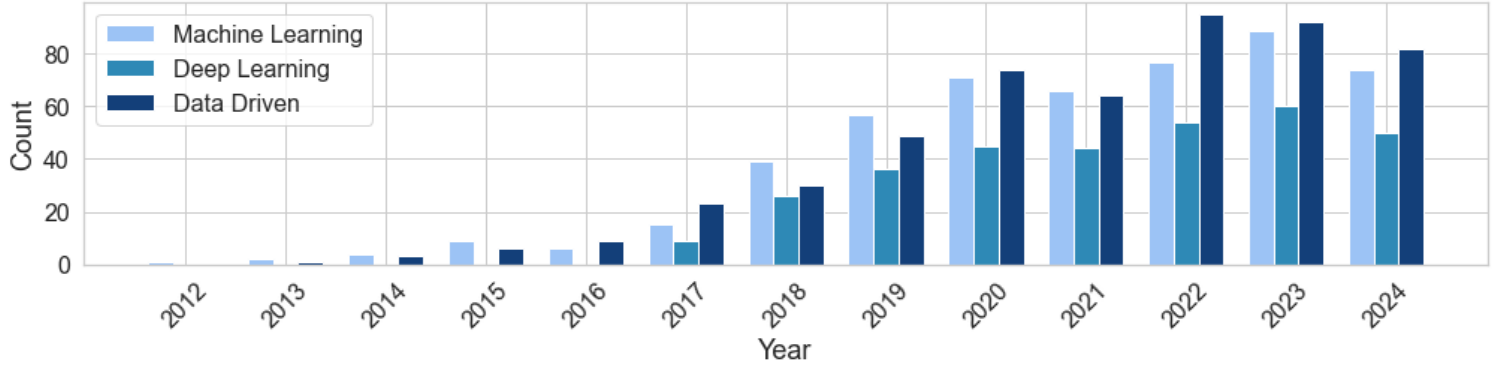}
  }
  \caption{
  The number of presentations from 2012-2024 APS-DFD meetings that contain ``machine learning," ``deep learning," and ``data driven" in their titles or abstracts (search performed on May 1, 2025). 
  }
\label{fig_count}
\end{figure}

To illustrate recent trends in the use of machine learning in fluid dynamics, we present in figure~\ref{fig_count} the number of the American Physical Society Division of Fluid Dynamics (APS-DFD) meeting abstracts that contain the term  ``data driven,'' ``machine learning,'' or ``deep learning'' \footnote{In contrast to these terms, it is interesting to note that the use of ``artificial intelligence'' has not been prevalent in fluid dynamics.}.  We observe that there has been a growing number of abstracts with machine learning/data-driven techniques being mentioned from 2012 to 2022.  It should be noted that there may be additional presentations that likely have discussed data-driven efforts without the aforementioned search words being listed.  What is also noteworthy is that from about 2022, there has been a plateau in the number of abstracts using those keywords.  We suspect that this plateau reflects the community no longer viewing data-driven approaches as something special, but rather, considering such techniques as part of the regular toolbox to study fluid dynamics.  As there is an increasing number of data-driven fluid dynamics workshops and conferences being held all over the globe, the impact of data-driven fluid dynamics is not plateauing but rising.

With the initial ramp-up phase of machine learning in fluid dynamics appearing to stabilize, data-driven fluid dynamics is coming to its transitional state from its initial growth stage to its next stage.  In some ways, the data-driven fluid dynamics community is entering the teenage years embarking on its soul-searching journey to determine how the subfield should shape itself as it further establishes its foundation and works together with the other subfields of fluid dynamics.  The purpose of this perspective article is to provide a critical assessment of the major outstanding issues in machine learning for fluid dynamics.  While we do not attempt to provide a review of the field or to predict future trends in this article, we hope the discussions provided herein stimulate and spark innovative ideas to support future machine learning research and educational activities in fluid dynamics.


\section{Discussion}

Let us offer a critical assessment of outstanding issues related to the use of machine learning in fluid dynamics.  
Although these issues may appear as hurdles at first glance, they could be areas of opportunity for future activities.  Even if they cannot be immediately addressed, we believe that being mindful of them can support individual studies and help shape community research directions.

\subsection{Formulation}

In the context of physical applications, including fluid mechanics, {\it machine learning} refers to the use of data-driven algorithms to uncover patterns, model complex dynamics, and approximate operators or functionals that may be difficult or impossible to express analytically.   A machine learning model $f_\theta$ is typically trained on simulation or experimental data to capture nonlinear relationships between the input features $\mathbf{x}_i$ (e.g., state variables, spatial coordinates, time, boundary conditions, or other problem-specific variables) and the output $\mathbf{y}_i$ (e.g., physical quantities composed of traditional or generic observables).  The machine learned model can be numerically found by solving a non-convex optimization problem of
\begin{equation}
\theta^* = \argmin_{\theta} \mathcal{L}(\theta) 
= \argmin_{\theta} \left[  \frac{1}{N}\sum_{i=1}^{N} \mathcal{L}_i(f_\theta(\mathbf{x}_i), \mathbf{y}_i) + \sum_{j} \lambda_j \mathcal{R}_j(\theta) \right].
\label{eq:ML}
\end{equation}
Here, the goal is to find the optimal set of model parameters $\theta^*$ that minimizes a loss function $\mathcal{L}(\theta)$.  The model parameters $\theta$ in modern machine learning are generally high-dimensional (e.g, weights of a neural network) \cite{Goodfellow-et-al-2016,Watt_Borhani_Katsaggelos_2020,Brunton_Kutz_2022}.  

One of the most critical steps in machine learning is the setup of this mathematical problem.   The loss function typically consists of two primary components: a data fidelity term that measures the discrepancy between model predictions $f_\theta(\mathbf{x}_i)$ and actual observations $\mathbf{y}_i$ across $N$ training samples, and regularization terms $\mathcal{R}_j(\theta)$ weighted by hyperparameters $\lambda_j$ to prevent overfitting \cite{Brunton_Kutz_2022}.  The parameter vector $\theta$ contains all the adjustable values in the model (e.g., weights and biases inside the neural network).  The resulting non-convex optimization landscape is typically navigated using variants of stochastic gradient descent that iteratively update parameters in the direction that reduces the loss function.  With this setting, a machine learning model can be used to represent a nonlinear operator, including a flow map, a closure model, or an input-output relation, enabling tasks such as reduced-order modeling, flow prediction, or control \cite{Brenner:PRF19,BNK2020,duraisamy2019turbulence}.

Generally, it is helpful to incorporate physical insights into the formulation of the problem to facilitate the machine-learning process and the interpretation of the results \cite{Brunton:AMC21,Koumoutsakos:DCE25}.  One way to achieve this is to add loss terms that penalize violations of the governing equations via the residuals of the relevant differential operators \cite{Raissi:JCP19,Karniadakis:NRP21,patel2024turbulence}, as part of the last term in Eq.~(\ref{eq:ML}).  The pros and cons of utilizing a specific machine learning algorithm and each loss function term should be considered carefully.  We caution that over-constraining the training objective is not advisable, as an appropriately compiled training data should allow a properly chosen machine learning algorithm to extract relevant information \cite{Sutton:bitter}.  One should also be mindful of whether a sufficient amount of training, validation, and test data is available to enable the machine-learning study to provide robust and generalizable findings.

Data-driven analysis should also be performed in steps with respect to the problem complexity.  It is often useful to test the desired machine-learning approach first for small-scale, canonical model problems before taking on the full-scale, complex fluid flow problems.  Although this workflow may seem obvious, the powerful capability of machine learning often oversteps it.   Such simple steps can, however, help with the verification of the approach and gain a deeper understanding of how the approach provides insights into the problem.  These steps can also identify the strengths and weaknesses of the selected approach in assessing the overall framework and may uncover possible improvements of the chosen model, mathematical problem setting, and the training dataset.

\subsection{Gaining Insights}

One of the biggest questions that the data-driven fluid dynamics community faces is whether the use of machine learning is actually deepening our understanding of fluid dynamics.  This seemingly simple question requires some thought.  First of all, what do we mean by understanding?   The term {\it understanding}, has different meanings depending on whether a person is trying to solve a practical problem or aiming to gain deep insights into a physical phenomenon.  These two objectives would lead to different ideas of insights.  In the case of solving a practical problem, information that helps solve the problem could be considered valuable insights.  However, such insights may not necessarily deepen our knowledge of physics.  In contrast, if the objective is to achieve enhanced knowledge of flow physics, we seek to relate the findings from machine learning to physical observations, knowledge, and theory at a deeper level while generalizing beyond the training data sets.  The latter objective calls for a more holistic approach and requires working with all four subfields of fluid dynamics.

We should also be cautious about the interpretability of nonlinear machine-learning methods.  It was only a few decades ago that proper orthogonal decomposition (POD) was considered an advanced (linear) technique that produced findings that were not interpretable by the fluid dynamics community.  However, the community now considers POD \cite{Berkooz:ARFM93,Taira_etal:AIAAJ17} as a mathematically rigorous data-driven technique that offers physically and mathematically interpretable results.  With the research community becoming familiar with the methods over time, what is considered ``interpretable" will evolve as the POD example reminds us.   This is also true for modern (nonlinear) machine-learning techniques for which well-experienced data-driven fluid dynamics researchers are now able to extract physical insights from latent variables \cite{Fukami:NC23}, which do not reside in the physical (ambient) space.  Such insights can be gained with proper experience in machine learning and by carefully relating the findings to fluid dynamics.  It is anticipated that many results produced from novel machine-learning techniques will be digested in a manner relatable to flow physics in the coming years, if not already.  As such, we expect the meaning of interpretability and the labeling of interpretable methods to actively evolve as the data-driven fluid dynamics field grows and matures in the future.

\subsection{Model Generalizability}

Given the diverse flow configurations, non-dimensional parameters, and a range of spatiotemporal flow characteristics we encounter in fluid dynamics, the generalizability of machine-learning models is a major challenge.  A truly generalizable model should perform well for a range of flow scenarios (e.g., from laminar to turbulent), while exhibiting robustness against variations in data fidelity, spatial resolution, and sensor noise.  Machine-learning models are, however, often reliable only within the regime for which testing conditions are similar to those seen in training.  We should be mindful that no single model can be optimal for all possible scenarios in the absence of prior knowledge \cite{wolpert1997no}.  This suggests a trade-off between model accuracy on a specific task and its ability to generalize across different flow conditions.

The validity of data-driven models is often discussed in the context of interpolation and extrapolation.  Modern machine-learning techniques may work outside the training regime if the models are able to capture existing trends across spatial scales or through nonlinear relationships.  This may appear as extrapolation in the traditional sense, but it generally amounts to interpolation in the machine learned coordinates.  Although caution is strongly advised, it is possible to observe pleasantly surprising performance even outside of the training data if the analysis is conducted appropriately \cite{Fukami:JFM24b}.  However, when the machine learning model fails outside the trained region over machine-learned coordinates, it does so spectacularly.  Understanding the transitional characteristics between interpolation and extrapolation requires a detailed analysis of the model in a machine-learned space that features the data distribution.  This is an open question in data-driven fluid dynamics that requires further examination, especially when identifying and modeling rare events.  

For applying trained models to unseen conditions, transfer learning can be useful.  This learning technique enables models trained with one particular flow problem to be reused and re-trained faster for another one.  Transfer learning has enabled the extension and fine tuning of machine learning models from lower Reynolds number flows for higher Reynolds number flows with reduced training efforts for flow estimation \cite{Guastoni:JFM21}, reinforcement learning based flow control \cite{Wang:JFM23}, and turbulence modeling \cite{Guan:JCP22}.  Another promising concept to extend the utility of existing models is the foundation model \cite{bommasani2021opportunities,Choi_foundation}.  This approach is often taken in image science, which fixes initial layers of a deep network to learn broadly applicable representations, while only fine-tuning the latter layers near the output for specific downstream tasks.  These general concepts may extend our current machine learning models to be applicable to a much wider range of fluid flow problems.

\subsection{Model Complexity}

Fundamental studies in the past have shown that functions or governing equations that describe physical phenomena are generally composed of simple, compact expressions.  In this vein, the best model should be parsimonious in its form for it to be interpretable and generalizable to describe the physics \cite{BNK2020,kutz2022parsimony,brunton2024promising,RichNAP95}.  
Basically, one should not use a complicated representation if there is a simple way to describe the phenomenon.  

As a general trend in any novel methods being used for fluid mechanics, initial efforts tend to be fairly simple in the way methods are utilized to solve problems.  Machine-learning techniques are no exception in how they have been introduced to the fluid dynamics community.  In recent years, however, the complexity of the machine-learning techniques being used for some studies has increased at a rate that has not been seen by our past data analysis techniques.  This is in part due to the machine-learning techniques' ability to hold a large number of parameters in the hidden layers and algorithms.  Tuning such a set of parameters used to be an enormous challenge for non-convex optimization problems (resulting from machine learning).  However, it is now becoming significantly easier with the advancement in software and hardware resources.  As such, there has been less hesitation for machine-learning practitioners to propose data-driven techniques that are surprisingly and unnecessarily complex, often due to a ``let's try and see if it works'' mentality.  For instance, some efforts have reported the use of machine-learning-based methods that require a number of tunable parameters far larger than what a computational fluid dynamics software would require to accurately simulate the flow to predict even the simplest flow behavior.

For a machine learning model, sparsity-promoting methods are capable of removing terms or network connections that are physically irrelevant or subdominant.  In many cases, they encourage interpretability of the returned results.  Moreover, when applied to the spatial domain, they can develop locally focused models instead of global ones.  These concepts have led to the identification of many sparse dynamical systems models \cite{Brunton:PNAS16} and partial differential equations \cite{Rudy:SA17} from flow field data.  

Without striving for simple or appropriate representation, machine-learning-based analysis may not be able to break away from being perceived as black-box models that as a conduit between the input and output data.  There are uses for such work, but they often would not yield deeper insights into the physics of fluid flows.

\subsection{Data Repositories and Code Sharing} 

As data plays an ever more important role in advancing science and engineering with machine learning, the utility of data should be carefully considered beyond a single or limited research use.  This means that training datasets should be compiled with downstream use for a large audience in mind.  
With machine learning algorithms generally being data-intensive, data as a resource should be shared across the community to advance the field of fluid dynamics.  Without such a shared resource, a researcher will be limited to a modest collection of data attainable by an individual research group, restricting machine learning research efforts.

There are three particularly important matters when compiling an archival data set.  First, to enable machine learning models to be as generalizable as possible across a range of conditions, the dataset must span a variety of physical phenomena.  Second, the dataset must be of high quality, faithfully capturing the intended flow physics without significant error or bias.  Third, the distribution of data should be carefully assessed to make sure sampling bias is minimized within an acceptable range \cite{Buolamwini,Efron_Hastie_2016}.  For example (while it may be unavoidable for now), there is a tendency for computational fluid dynamics data to be associated with lower Reynolds numbers compared to experimental measurements.

The fluid flow data repositories hosted by Johns Hopkins University for canonical turbulent flow data sets \cite{JHTDB1,JHTDB2} have supported the development of analysis and modeling of turbulent flows with lasting impact. There are also repositories made available in recent years with unsteady flow data for machine learning applications in mind \cite{Towne:AIAAJ23, chung2023turbulence}.  As machine learning models can learn flow physics from not only a limited number of cases but a large number of flows across combinations of parameters, we require a much larger data repository comprising a sizable ensemble of cases \cite{chung2023turbulence}.  This means that such a dataset would have a large number of spatial grid points, temporal snapshots, and flow cases, making a necessary data library to be of a completely different size from what we are accustomed to hosting by a single research group or center.  

Consequently, one of the major issues is the hosting of such a data repository.  While most research support is provided for research projects and equipment, rarely do we see support for hosting a data server with dedicated staff.  This matter needs serious attention to enable researchers to freely share datasets and advance machine learning research efforts.  Moreover, it may require a change in our perception to consider flow data repositories as important archival resources that may require libraries or national/international research centers to curate important spatiotemporally resolved flow field data across a range of parameters for various flows.  In fact, our community is in need of a discussion to enhance findability, accessibility, interoperability, and reusability (FAIR) \cite{FAIR16} of training data sets for machine learning in fluid dynamics.  We can learn from other areas of research, such as the astronomy and atmospheric science communities, that have a communal data repository.  Once the fluid dynamics community establishes guidelines on data sharing setups and formats, we will be able to further accelerate the research progress in machine learning and develop generalizable and robust models.  It may also invite new data formats for analysis beyond probe data and snapshots \cite{Anantharaman:TCFD23}.  Such a data repository will bring benefit to the entire fluid dynamics community beyond the machine learning groups and the scientific community at large.  

In support of FAIR, it is important to share the codes for machine-learning analysis in the community.
There has been encouraging progress by some journals and research groups, making their machine-learning codes and datasets publicly available through online platforms such as GitHub and Zenodo.
Such efforts can enhance not only the reproducibility of the results but also collaboration across the world. 
To further support this promising trend, it would be necessary to address several remaining challenges, such as a lack of standardization in code structure and documentation, likely causing difficulties in reproducing results and extending the code to their own data sets.
In addition to these technical matters, the community needs to culturally shift toward a state of transparent research practices \cite{NMIeditorial25}.

Let us also mention an area that requires some attention related to the coding effort.  There are well-established guidelines on verification, validation, and uncertainty qualification for computational and experimental fluid dynamics \cite{AIAA_VnV,tropea2007springer,Smith14}.
Such procedures ensure that studies performed in these fields provide reliable insights. 
Unfortunately, for many machine-learning efforts, there does not appear to be a similar level of rigor that establishes the architectures of the models and ensures the fidelity and performance of their findings \cite{McGreivy:NMI24,Koumoutsakos:DCE25}.  
This is not to say such processes do not exist in the machine-learning community.  
However, they are often overlooked. 
Furthermore, given the complexities of machine learning, it would be ideal to have the ability to estimate its necessary computational and memory resources in a manner analogous to how scaling is performed to determine the required grid resolutions across Reynolds numbers in computational fluid dynamics analysis.
This current state of the matter urgently calls for some guidelines to be developed such that the robustness and fidelity of machine-learning techniques can be guaranteed in future efforts.
With proper verification and validation performed and limitations identified, higher credibility can be attributed to machine-learning models.  

While sharing resources should be encouraged, we should remember that collecting high-quality data is an endeavor of its own.  Data-driven fluid dynamics is reliant on experts in experimental and computational fluid dynamics to collect the data.  Proper credit always needs to be given to the data creators/collectors.

\subsection{Education}

Given the enormous progress and excitement seen in the general machine learning community, it is perhaps understandable to have high expectations that machine learning can serve as a breakthrough technology to solve some of the most challenging problems in fluid dynamics. However, we must have healthy expectations of what can be achieved by machine learning.  It is important to recognize that not all developments in general machine learning can be easily transferred to the fluid dynamics community due to the difference in the research objectives and the characteristics of flow physics.  Machine learning practitioners in fluid dynamics should also have a deep understanding of the flow physics, as incorporating prior knowledge about the flow can improve the machine learning process and facilitate the extraction of insights from data.  For undergraduate and graduate education, a class on data-driven fluid dynamics should be developed in concert with theoretical, computational, and experimental fluid dynamics courses.  Furthermore, exposing students to machine learning in their early stages of fluid mechanics education through homework assignments and projects that involve data-driven analysis may strengthen their ability to use such tools to a broader class of problems and prepare them for proper usage of machine learning analysis in fluid dynamics.

There is a new type of problem that our community is facing in terms of the need for machine learning education.  As the younger generation enters the research community already with some familiarity with machine learning, it can be challenging even for established researchers to keep up with the younger minds.  Many engineering and science students now enter graduate school with machine learning experience.  This raises a question of not only how the community should teach and advise the incoming research population, but also how we should train the pre-machine learning research population. This generational gap in machine learning training needs to be filled to establish a balanced perspective on fluid dynamics and maintain healthy expectations on what machine learning can achieve in research and development.  For these reasons, educational opportunities should be made available to the wider research population at different levels of familiarity with machine learning.  

In response to these points, data-driven fluid dynamics conferences and workshops should involve a diverse set of stakeholders, including those from all four subfields of fluid dynamics working in academia, industry, and government.  Moreover, workshops aimed at non-practitioners are equally important as those targeted at spreading the state of the art of machine learning.  By doing so, we will be able to achieve the integration of machine learning techniques in various studies of fluid dynamics, when appropriate, and advance the overall field.  Essentially, we will be able to blend the four subfields of fluid dynamics seamlessly across generations.  
Moreover, fundamental and applied machine learning research in fluid dynamics must work hand-in-hand. 
Only with a fundamental understanding of how machine learning can make a difference in fluid mechanics, the swift transfer of machine learning technology from basic science to practical applications can be enabled \cite{Brunton:AIAAJ21,Tran:CE24}.    

At last, machine learning practitioners in fluid dynamics should have good knowledge of computational and experimental fluid dynamics to understand what constitutes good data.  It is also important to appreciate how uncertainties and biases can creep into the training dataset.  As machine learning intimately relies on the quality of the training data, the data-driven fluid dynamics community must work closely with the experts who collect high-quality data from simulations and experiments.  Blending these subfields of fluid dynamics will advance the reliability of future machine learned models and enhance the insights gained from their learning process.


\section{Remarks}

Data-driven fluid dynamics has undergone tremendous growth over the past decade and has become recognized as a powerful fourth subfield in fluid dynamics, complementing theoretical, experimental, and computational fluid dynamics.  In fact, the usage of data-driven techniques has become so widespread in fluid dynamics research that it is no longer special to see it in conference presentations or archival articles.  As the field of data-driven fluid dynamics aims to advance itself further, it requires careful examination of some outstanding issues.  This perspective article calls attention to machine learning formulation, gaining insights, generalizing models, considering model complexity, sharing data and codes, and education.  It is our sincere hope that this article stimulates future discussions and invites innovative ideas.

There remain many challenging problems in fluid dynamics that involve multi-physics and unknown mechanisms.  These problems may not have an established governing equation, for which machine learning may serve as a powerful tool to make headway.  It may also perform well for problems with strong nonlinearity that traditional techniques have difficulty analyzing.  There are fluid dynamics problems that do not have the luxury of having sufficient data due to challenges associated with conducting a large number of experiments or simulations.  Machine learning efforts will have to adapt their techniques to sample and leverage data in a smart manner.  In such a case, incorporating prior knowledge about the physics may facilitate the learning process even from a minimal amount of data \cite{Fukami:JFM24c}.  Moreover, there is growing interest in applying generative models, including large language models, diffusion models, and other foundation models \cite{CoTLLM, Bodnar:Nature25}, to fluid mechanics problems. These approaches can employ chain-of-thought reasoning \cite{CoTLLM} to explicitly trace the logical steps underlying fluid flow analysis and prediction. By making the reasoning process transparent, these techniques offer potential for both deepening our mechanistic understanding of fluid phenomena and improving predictive accuracy across diverse flow regimes \cite{Bodnar:Nature25}.

The development of machine learning techniques has been closely tied to hardware advancements.  The availability of powerful GPUs has catapulted the recent movement in deep learning.  With the GPU architecture enabling large-scale nonlinear optimization problems to be numerically solved, deep learning methods and techniques evolving around them have been enjoying the benefits.  The current algorithmic trends will evolve in the future as novel computing architectures elicit innovations in next-generation machine learning approaches, including those associated with quantum computing \cite{Biamonte:Nature17,Jaksch:AIAAJ23}.  As CPUs and GPUs have influenced computational science and machine learning, the future of fluid dynamics will likely see a transformation with advancements in future computing platforms.

Before closing, let us briefly comment on publishing works on data-driven fluid dynamics.  The community should maintain the same high standards of publication for manuscripts on the topic of theoretical, experimental, computational, and data-driven fluid dynamics, or combinations thereof.  This means that papers in data-driven fluid dynamics should deepen our understanding of flow physics or propose novel methodologies that can aid discoveries in fluid dynamics \cite{BrennerKoumoutsakos:PRF21}.  For the latter type of work, Physical Review Fluids houses a topical section entitled {\it Methods: New Experiments, Algorithms, and Theory} (NEAT) \cite{PRFluids_NEAT}, which can serve as a home to breakthrough data-driven fluid dynamics papers.  As our research community goes through this transitional period for machine learning in fluid dynamics, it is very exciting to imagine the kinds of innovative papers we will see in the coming years. 

The vast advancements made by data-driven fluid dynamics have been nothing short of incredible.  With the aforementioned outstanding problems addressed in the coming years, one can only imagine how much progress the community will be able to make.  The intellectual investments we make in the data-driven fluid dynamics during the current transitional state will form the basis of how machine learning will expand the future horizon of fluid dynamics in concert with theoretical, experimental, and computational efforts.


\section*{acknowledgments}
   We thank Jeff D.~Eldredge, Alec J.~Linot, Beverley J.~McKeon, Douglas R.~Smith, and Jonathan Tran for their valuable comments on the draft version of this article.  
   KT acknowledges support from the Vannevar Bush Faculty Fellowship (N00014-22-1-2798), the Air Force Office of Scientific Research (FA9550-21-1-0178, FA2386-25-1-1003), and the Army Research Office (W911NF-24-1-0213). 
   GR acknowledges support from the UKRI AI for Net Zero grant (EP/Y005619/1).
   KF acknowledges support from the JSPS KAKENHI Grant Number JP25K23418.


\bibliographystyle{unsrt}  
\bibliography{Taira_refs} 

\end{document}